\documentclass[final,3p,times]{elsarticle}

\usepackage{amssymb}
\usepackage{amsmath}

\begin{document}

\begin{frontmatter}



\title{Chiral symmetry breaking, chiral partners, and the $K_1$ and $K^*$ in medium}


\author{Su Houng Lee} 
\ead{suhoung@yonsei.ac.kr}

\affiliation{organization={Department of Physics, Yonsei University},
            addressline={50 Yonseiro}, 
            city={Seoul},
            postcode={03722}, 
            country={Korea}}

\begin{abstract}
We clarify the concept of chiral partners. For a vector meson, the isospin-zero and hypercharge-zero state in the flavor octet mixes with the flavor singlet state. Since the flavor singlet vector meson does not have a chiral partner, the mixed $\omega$ and $\phi$  mesons will not have chiral partners. This means that even when chiral symmetry is restored, these mesons will not become degenerate with their corresponding parity partners. 
On the other hand, the $K_1$ and $K^*$ 
mesons are chiral partners, and both have widths smaller than 100 MeV. Therefore, we emphasize that studying these mesons in environments where chiral symmetry is restored is particularly important for understanding the effect of chiral symmetry restoration on chiral partners and their masses.
\end{abstract}




\end{frontmatter}



\section{Introduction}
\label{sec1}

The masses of the $u$ and $d$ quarks are both less than 10 MeV$/c^2$, but the mass of a proton, which is composed of three quarks, is 938 MeV$/c^2$. Such a large mass difference between a particle and its constituents is also true for any hadron, such as the vector meson which is composed two quarks but has a mass of around 780 MeV$/c^2$. The situation is in sharp contrast to what is typically known of a composite particle such as the hydrogen atom, the mass of which can be understood as the sum of the masses of its constituents, namely the proton and electron, and a small binding energy. The puzzle is related to chiral symmetry breaking and confining phenomena in Quantum Chromodynamics, which is known to be the underlying theory for strong interaction.

In ref \cite{Hatsuda:1991ez}, we showed that the mass of the vector meson will change in nuclear matter due to partial chiral symmetry restoration .  The work attracted great attention as such a change can be observed from a heavy nucleus target experiment, which will provide a link to understanding the origin of hadron masses\cite{Hatsuda:1985eb,Brown:1991kk,Leupold:2009kz,Gubler:2016djf,Song:2018plu}.  Unfortunately, while several experiments tried to observe the vector mass shift through electromagnetic signals from heavy ion collision and/or nuclear target experiments\cite{Hayano:2008vn}, no solid confirmation of $\rho$ meson mass shft has been seen so far except for increased width\cite{Rapp:1999ej}.  
One important drawback of measuring the 
$\rho$ meson is that, while it contributes dominantly to the electromagnetic signal in the intermediate energy region, its large vacuum width, along with additional broadening in the medium, will make any realistic mass shift measurement impossible\cite{Leupold:2009kz}. 
Furthermore, in addition to chiral symmetry-breaking effects, the $\rho$ meson mass has contributions from other effects so it is hard to identify the origin of the hadron mass as generated in the vacuum in terms of underlying QCD degrees of freedom. 
To isolate the effects of chiral symmetry breaking, one can study the mass difference between chiral partners. However, the chiral partner of the $\rho$ meson, the $a_1$ meson, has an even larger vacuum width, making any realistic measurement in the medium impossible.

The recent efforts by the E16 experiment \cite{JPARC:2023quf,Aoki:2023qgl} to measure the $\phi$ meson mass through its electromagnetic decay in a nuclear target with higher statistics, compared to the previous KEK experiment\cite{KEK-PS-E325:2005wbm}, are very promising, particularly due to the narrow vacuum width of the $\phi$ meson. Furthermore, the E-88 experiment\cite{Sako} will measure the $\phi$ meson mass shift through the 
$K^+K^-$ decay. Both experiments are promising and could provide the first explicit measurement of a mass shift.

On the other hand, $\phi$ does not have an exact chiral partner per se, as will be discussed in the next section.  In the following, we will discuss the meaning of chiral partners and why $K_1,K^*$ are good candidates of chiral partners that can be realistically measured\cite{Lee:2019tvt}. 

\section{Flavor SU(3) symmetry and octet singlet mixing}

\begin{figure}[t]
\centering
\includegraphics[scale=0.5]{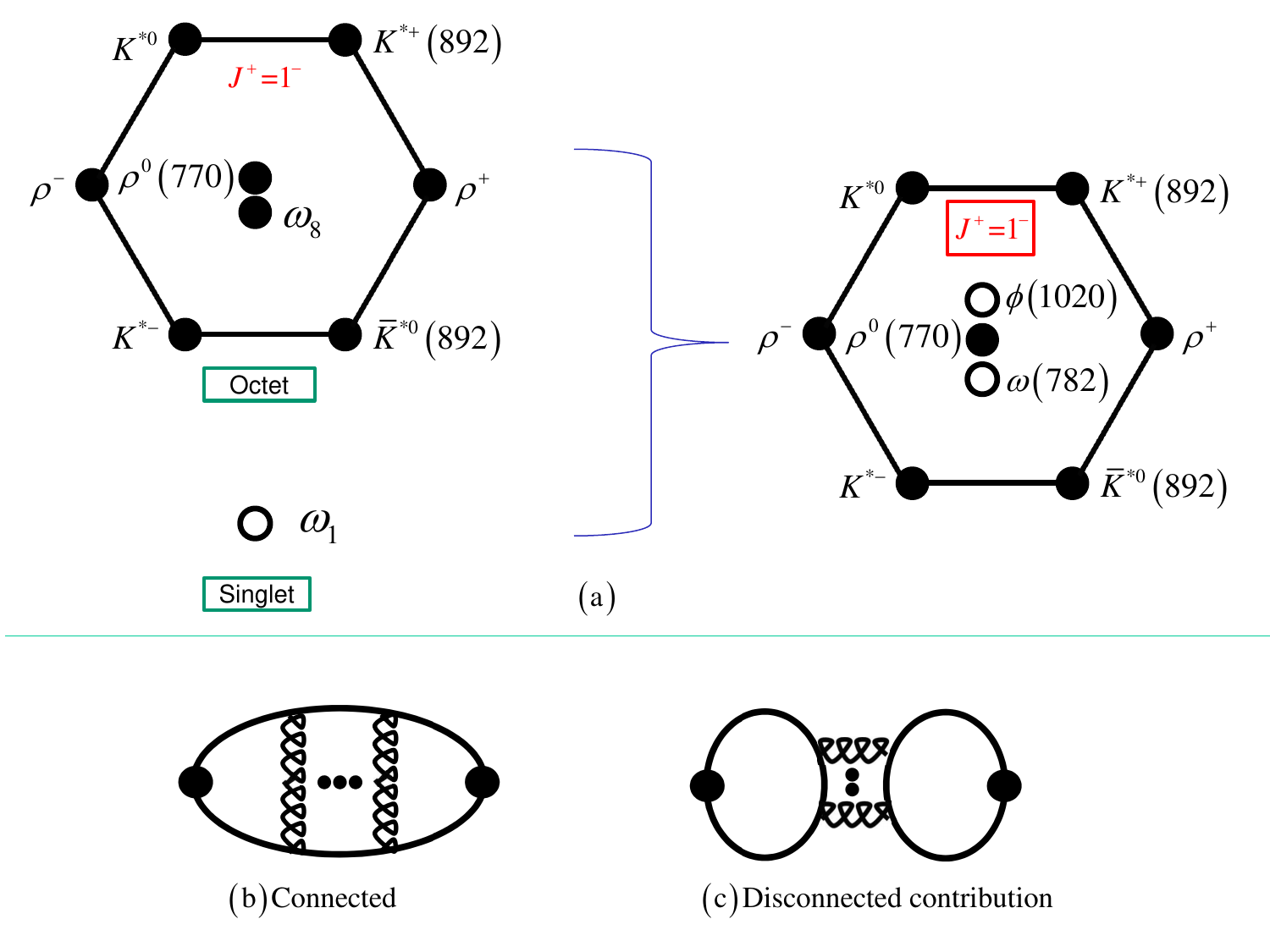}
\caption{Ideal mixing for the vector meson}\label{fig-1}
\end{figure}

If all the light quark masses were equal, the meson spectrum would form an irreducible representation of flavor SU(3). For meson states, this would include a flavor singlet and octet. However, because the strange quark mass is significantly larger than the up and down quark masses, the two states with zero isospin and hypercharge in the singlet and octet representations will mix.  
How much the octet and singlet mix depends on the nature of the particles. For the pseudoscalar mesons, the mixing is small, while for the vector mesons, it is almost ideal.  The situation is depicted in Fig.~\ref{fig-1}-(a).  Here the fields are as follows.
\begin{align}
\begin{cases} 
\omega_1=\frac{1}{\sqrt{3}}(\bar{u}u+\bar{d}d+\bar{s}s )  \cr
\omega_8=\frac{1}{\sqrt{6}}(\bar{u}u+\bar{d}d-2\bar{s}s ) \end{cases} \stackrel{m_s>>m_u,m_d}{\longrightarrow}
\begin{cases} 
\omega=\frac{1}{\sqrt{2}}(\bar{u}u+\bar{d}d)  \cr
\phi=(\bar{s}s ) \end{cases} .
\nonumber
\end{align}

The origin of ideal mixing can be understood using correlation functions.  For that purpose, consider two currents $\omega_1=\bar{q}q+\bar{s}s$ and $\omega_3=\bar{q}q -\bar{s}s$. Here $q$ represents the light quark and $s$ represents the strange quark. We limit the discussion to SU(2), but the argument can be easily generalized to SU(3).
Consider the two-dimensional correlation function composed of $\langle (\omega_i )(\omega_j) \rangle $, where $i,j=1,3$. The matrix can be written as follows.
\begin{equation}
\begin{pmatrix} 
a +D &  \Delta  \\
\Delta   &  a -D  
\end{pmatrix},
\end{equation}
where 
\begin{eqnarray}
a &=&  \langle (\bar{q}q)(\bar{q}q) + (\bar{s}s)(\bar{s}s) \rangle, \\
\Delta &=&  \langle (\bar{q}q)(\bar{q}q) - (\bar{s}s)(\bar{s}s) \rangle, \\
D  &=& \langle 2 (\bar{q}q)(\bar{s}s) \rangle. 
\end{eqnarray}
$\Delta$ determines the magnitude of SU(3) symmetry breaking, while $D$ has contributions from disconnected quark diagrams only.   The connected and disconnected are diagrammatically represented in Fig.~\ref{fig-1}-(b) and (c), respectively.  
The magnitude of mixing between the 1 and 2 components is determined by the relative strength between $\Delta$ and $D$. 
The two extreme limits are as follows

\begin{enumerate}
    \item $\Delta=0$ corresponds to the SU(3) symmetric limit.  In this limit, there is no mixing as expected.
\begin{equation}
\begin{pmatrix} 
a +D &  \Delta  \\
\Delta   &  a -D  
\end{pmatrix} \rightarrow
\begin{pmatrix} 
\langle (\bar{q}q)(\bar{q}q) + (\bar{s}s)(\bar{s}s) +2 
\langle (\bar{q}q) (\bar{s}s) \rangle
\rangle &  0  \\
0  &  \langle (\bar{q}q)(\bar{q}q) + (\bar{s}s)(\bar{s}s) -2 
\langle (\bar{q}q) (\bar{s}s) \rangle
\rangle
\end{pmatrix}.
\end{equation}    

    \item $D=0$ corresponds to the limit where the quark disconnect contribution is zero.  In this case, the matrix can be diagonalized with eigenvalues equal to $a\pm \Delta$, which corresponds to the ideal limit. 
\begin{equation}
\begin{pmatrix} 
a +D &  \Delta  \\
\Delta   &  a -D  
\end{pmatrix} \rightarrow
\begin{pmatrix} 
\langle 2 (\bar{q}q)(\bar{q}q) \rangle  &  0  \\
0  &  \langle 2(\bar{s}s)(\bar{s}s) \rangle
\end{pmatrix}.
\end{equation}

\item If both $\Delta$ and $D$ are non-zero, the mixing angle depends on the magnitude of the terms. If $D/\Delta$ is small, the mixing angle is close to the ideal mixing limit. On the other hand, if $D/\Delta$ is large, the mixing angle is small. The former case applies to the vector mesons.
\end{enumerate}

\section{Chiral Partners}

\subsection{Chiral symmetry}

The QCD Lagrangian has SU(3)$_{\rm L} \times$ SU(3)$_{\rm R}$ symmetry when the three flavors are massless. The symmetry can be represented as invariance under the following transformations for quarks.
\begin{eqnarray}
q_{\rm R,L} \rightarrow \exp\bigg( i \vec{ \theta}_{\rm R,L}(1 \pm \gamma^5) \bigg)q_{\rm R,L},
\end{eqnarray}
where $\vec{\theta}=\theta^a \lambda^a$ is flavor valued.

In the physical QCD vacuum, the chiral symmetry is spontaneously broken.
\begin{align}
    {\rm SU(3)_R} \times {\rm SU(3)_L} \rightarrow {\rm SU(3)_V}.
\end{align}
This means that the ground state and the physical states will not be invariant under the following quark transformation.
\begin{eqnarray}
    q \rightarrow \exp\bigg( i \vec{ \theta} \gamma^5 \bigg)q. \label{eq:chiral-t} 
\end{eqnarray}

Let us consider the transformation of quark bilinears under Eq.~(\ref{eq:chiral-t}) when ${\bf \theta}$ is small. 
\begin{eqnarray}
\bar{q} \Gamma q \rightarrow
\bar{q} \Gamma q +
 \bar{q} 
\bigg( i \vec{ \theta} \gamma^5 \Gamma+ \Gamma i \vec{ \theta} \gamma^5  \bigg)q.
\end{eqnarray}
Here $\Gamma$ can involve both Dirac and flavor matrices.
Specific examples are as follows.
\begin{align}
\begin{cases} 
\bar{q} q  & \rightarrow 
\bar{q} \Gamma q + 2
\bar{q} 
\bigg( i \vec{ \theta} \gamma^5  \bigg)q,  \cr
\bar{q} \gamma^\mu \tau^a  q & \rightarrow 
\bar{q} \gamma^\mu \tau^a q +
\bar{q} 
i \gamma^5 \gamma^\mu \bigg[\vec{ \theta} , \tau^a \bigg] q, \cr
\bar{q} \gamma^\mu  q & \rightarrow 
\bar{q} \gamma^\mu  q +
\bar{q} .
i \gamma^5 \gamma^\mu \bigg[\vec{ \theta} , 1 \bigg] q = \bar{q} \gamma^\mu  q
\end{cases} .
\end{align}
Therefore, while the vector mesons in the flavor octet representation transform into the axial vector mesons within the octet representation, the flavor singlet vector and axial vector mesons do not transform into each other.

\subsection{Chiral partners}

Therefore, when flavor SU(3) symmetry is exact, the vector mesons in the flavor octet representation are chiral partners of the axial vector mesons in the corresponding flavor octet representation. In contrast, the flavor singlet vector and axial vector mesons are not chiral partners.

In the real world where the SU(3) flavor symmetry is broken, and the $\phi$ and $\omega $ mesons are ideally mixed states between flavor octet and singlet components. As a result, they contain components that do not have chiral counterparts and, therefore, cannot be considered chiral partners with their corresponding axial counterparts.  One can call these parity partners. 
One can still discuss the chiral transformation in Eq.~(\ref{eq:chiral-t}), where $\theta$ is SU(3)-valued, and the flavor octet vector mesons are related to the flavor octet axial vector mesons through chiral transformations. However, since the explicit breaking of SU(3) symmetry is significant in the strangeness direction, it is more appropriate to focus on the chiral transformation limited to the light quark sector, where $\theta$ is SU(2)-valued. In this limit, the $a_1$ is the chiral partner of the $\rho$, and the $K_1$ is the chiral partner of the $K^*$.

\begin{figure}[t]
\centering
\includegraphics[scale=0.5]{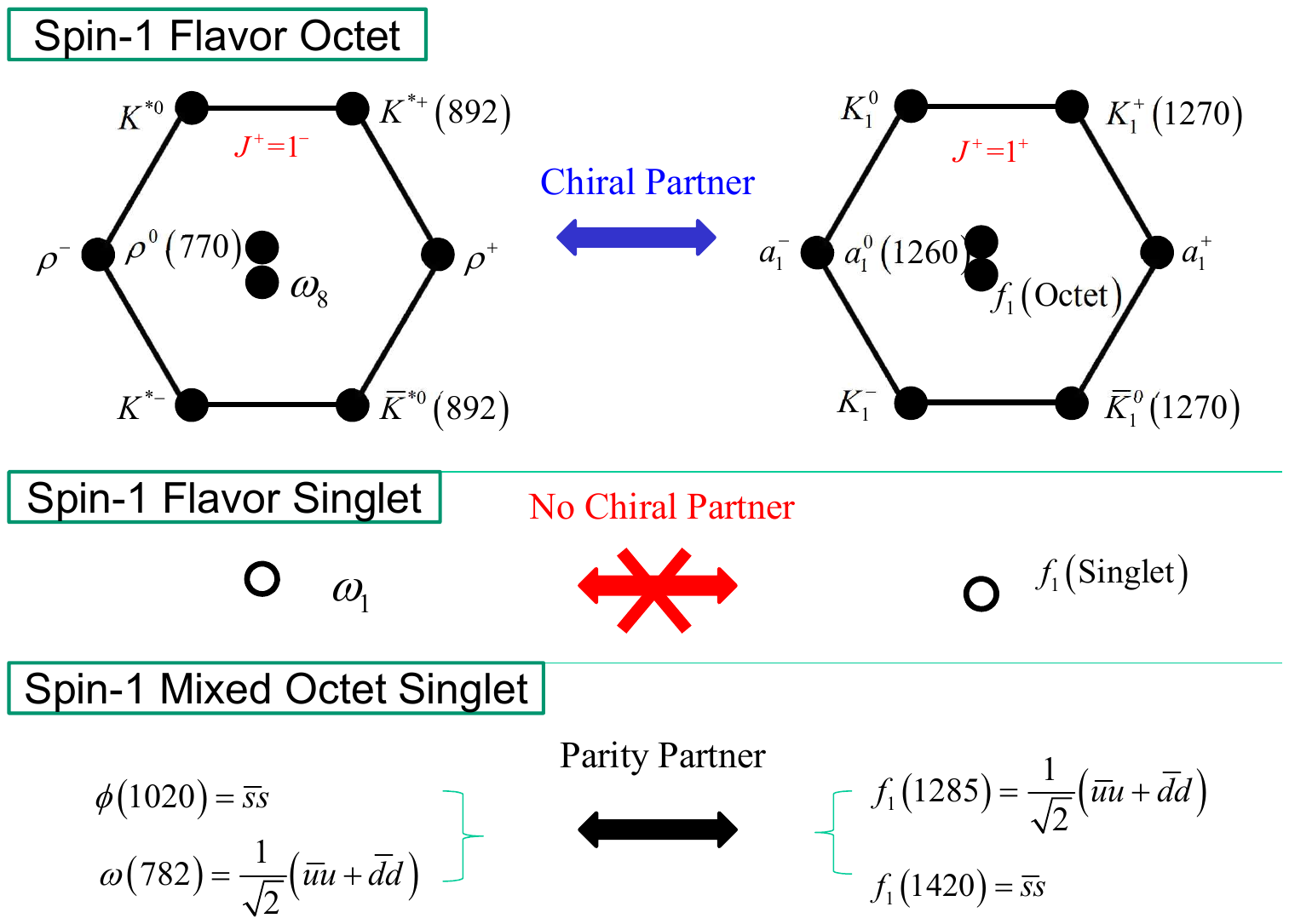}
\caption{Chiral partners and Parity partners.}\label{fig-2}
\end{figure}

\section{Chiral symmetry restoration}

The broken chiral symmetry is expected to be restored at high temperature and/or density.  
In general, in addition to chiral symmetry restoration, other effects will occur at finite temperature and/or density.  Hence, one can not isolate the effects of chiral symmetry restoration in the mass changes occurring in the individual hadrons.  
On the other hand, if one studies the chiral partners, all effects other than chiral symmetry breaking will cancel out and the masses of chiral partners will become degenerate.  

This means that in the SU(3) symmetric case, all the vector meson masses in the octet will become degenerate with the axial vector meson masses in the same representation.  On the other hand, the singlets will become non-degenerate.  

\begin{figure}[t]
\centering
\includegraphics[scale=0.5]{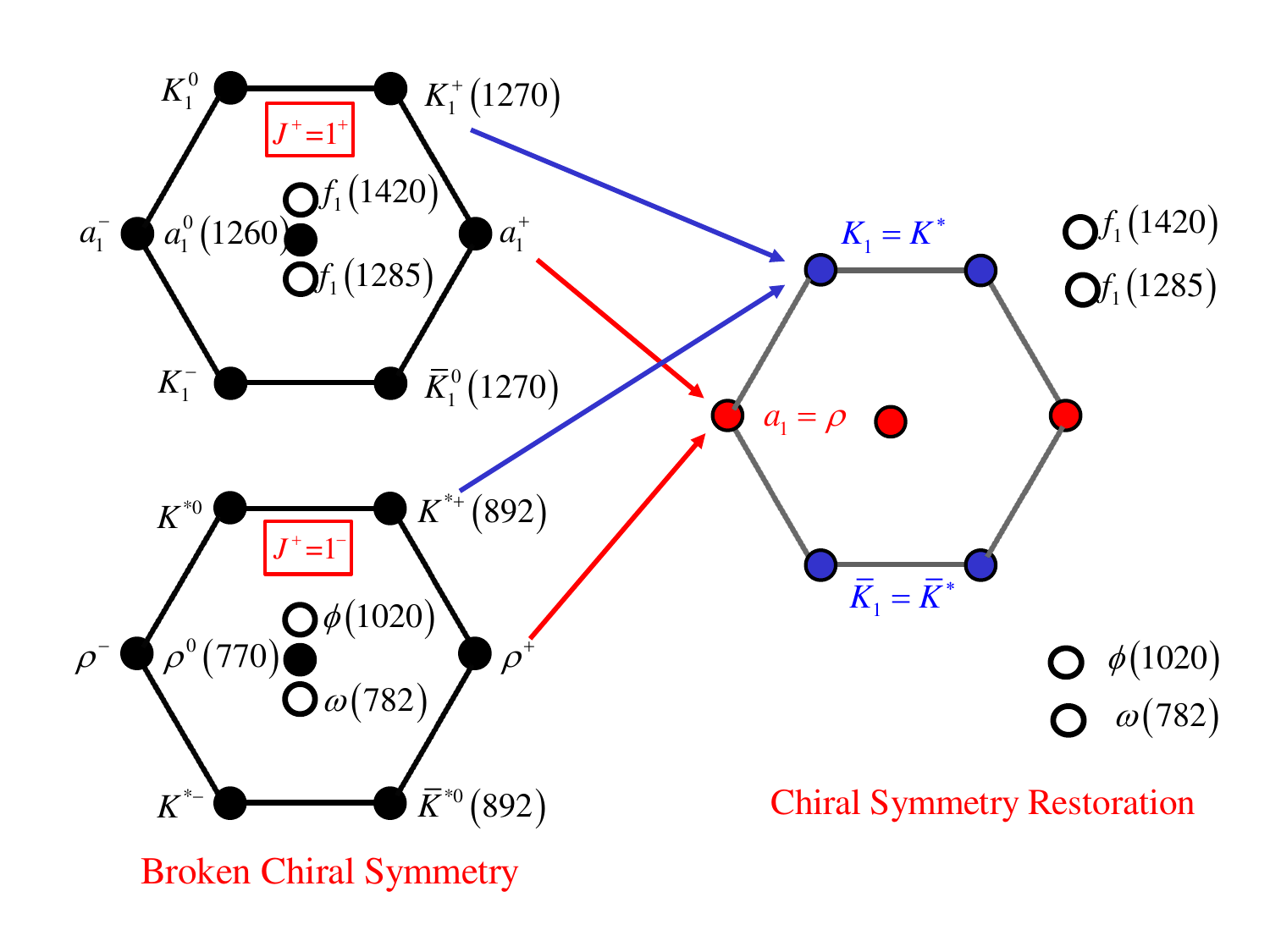}
\caption{Figure Caption}\label{fig2}
\end{figure}

In real QCD, where the SU(3) symmetry is broken in the hypercharge direction, one can expect the masses to become degenerate in the SU(2) subspaces as shown Fig.~\ref{fig2}.  
The $K^*$ and $\rho$ mesons will become degenerate with the $K_1$ and $a_1$ mesons, respectively.  On the other hand, $\phi$ and $\omega$ mesons will not become degenerate with their corresponding parity partners: namely the $f_1(1420)$ and $f_1(1285)$, respectively.   This is because, $\phi$ and $\omega$ mesons are mixed flavor octet and singlet states, the latter of which does not have a chiral partner.

\subsection{Chiral partner from Operator Product Expansion}

The relation between chiral partners can be well understood in terms of the operator product expansion (OPE)\cite{Lee:2023ofg}. To illustrate this, let us consider the $K_1-K^*$ system. 
We choose the currents to be $J^{K^*}_\mu =\bar{u} \gamma_\mu s$ and $J^{K_1}_\mu =\bar{u}  \gamma_\mu \gamma^5 s$ and study the correlation function in the momentum space.  The corresponding  dimension-6  scalar 4-quark operators in $-\Pi^\mu_\mu(Q^2)/3$ to leading order are given below. 
\begin{eqnarray}
\Pi^{K^*} & = & - 
\frac{2 \pi \alpha_s}{Q^6} \bigg(
\langle ( \bar{u} \gamma_\mu \gamma^5 \lambda^a s )( \bar{s} \gamma_\mu \gamma^5 \lambda^a u ) \rangle
+ \frac{1}{9} 
\langle ( \bar{s} \gamma_\mu \lambda^a  s +\bar{u} \gamma_\mu \lambda^a  u )
( \bar{q} \gamma_\mu \lambda^a  q ) \rangle \bigg) ,
\nonumber \\
\Pi^{K_1} & =  & -
\frac{2 \pi \alpha_s}{Q^6}  \bigg(
\langle ( \bar{u} \gamma_\mu \lambda^a s )( \bar{s} \gamma_\mu  \lambda^a u ) \rangle
+ \frac{1}{9} 
\langle ( \bar{s} \gamma_\mu \lambda^a  s +\bar{u} \gamma_\mu \lambda^a  u )
( \bar{q} \gamma_\mu \lambda^a  q ) \rangle \bigg).
\label{ks-k1}
\end{eqnarray}
The first terms in both OPEs are different but the second terms are identical.  All these operators can be decomposed in terms of chiral symmetry breaking and symmetric pieces\cite{Kim:2020zae,Kim:2021xyp}.   

When the difference between the chiral partners is taken, only the first terms from the correlation functions contribute.  This difference can be written as follows.
\begin{eqnarray}
\Pi^{K^*}-\Pi^{K_1} & = & 
\frac{2 \pi \alpha_s}{Q^6} \bigg(
\langle  \bar{s} \gamma_\mu  \bigg[ S_u(x,0)- i\gamma^5 S_u(x,0) i \gamma^5 \bigg] \lambda^a \gamma^\mu u ) \rangle +( s \leftrightarrow u) \bigg) .
\label{ks-k1-2}
\end{eqnarray}
Here, the term inside the large square bracket is the difference between the quark propagator and its chiral transformed counterpart, which is the chiral order parameter appearing in the quark condensate.  
\begin{equation} 
\begin{split} 
\langle \bar{q} q \rangle &= - \lim_{x \rightarrow 0}  \langle {\rm Tr} \frac{1}{2} \bigg[ S_u(x,0)- i\gamma^5 S_u(x,0) i \gamma^5 \bigg] 
\rangle = -\pi \langle \rho(0) \rangle, 
\label{bc-rel}
\end{split} 
\end{equation}
where $\rho(0)$ is the zero mode density, a formula derived by Casher and Banks \cite{BC}. 
Hence, one notes that the difference between the correlation functions of chiral partners is a chiral order parameter.

\subsection{Difference between non-Chiral partners}

If one takes the difference between parity partners or between currents composed of a mixture between the octet and singlet representations, the difference has contributions not related to chiral order parameter. 
To illustrate this, let us consider the $\phi-f_1(1420)$ system. 
We choose the currents to be $J^{\phi}_\mu =\bar{s} \gamma_\mu s$ and $f_1(1420)_\mu =\bar{s}  \gamma_\mu \gamma^5 s$, the corresponding  dimension-6  scalar 4-quark operators to leading order are given below. 
\begin{eqnarray}
\Pi^{\phi} & = & - 
\frac{2 \pi \alpha_s}{Q^6} \bigg(
\langle ( \bar{s} \gamma_\mu \gamma^5 \lambda^a s )( \bar{s} \gamma_\mu \gamma^5 \lambda^a s ) \rangle
+ \frac{2}{9} 
\langle ( \bar{s} \gamma_\mu \lambda^a  s  )
( \bar{q} \gamma_\mu \lambda^a  q ) \rangle \bigg) ,
\nonumber \\
\Pi^{f_1(1420)} & =  & -
\frac{2 \pi \alpha_s}{Q^6}  \bigg(
\langle ( \bar{u} \gamma_\mu \lambda^a s )( \bar{s} \gamma_\mu  \lambda^a u ) \rangle
+ \frac{2}{9} 
\langle ( \bar{s} \gamma_\mu \lambda^a  s  )
( \bar{q} \gamma_\mu \lambda^a  q ) \rangle \bigg).
\label{phi-f1}. 
\end{eqnarray}

Again, the difference comes the first terms.   This difference can be written as follows.
\begin{eqnarray}
\Pi^{\phi}-\Pi^{f_1(1420)} & = & 
\frac{2 \pi \alpha_s}{Q^6} \bigg(
\langle  \bar{s} \gamma_\mu  \bigg[ S_s(x,0)- i\gamma^5 S_s(x,0) i \gamma^5 \bigg] \lambda^a \gamma^\mu s ) \rangle +[{\rm Disconnected} ] \bigg),
\label{phi-f1-2}
\end{eqnarray}
where $S_s$ is the s-quark propagator, so that the term inside the large square bracket is now related to both explicit and spontaneous chiral symmetry breaking effects due to the large and non-vanishing strangequark mass.  
However, there is an additional contribution from disconnected diagrams, which can be written as
\begin{eqnarray}
[{\rm Disconnected } ] & = & 
\frac{2 \pi \alpha_s}{Q^6} \bigg( {\rm Tr}\big[S_s(x,x)\gamma_\mu \lambda^a \big] 
{\rm Tr}\big[S_s(0,0)\gamma_\mu \lambda^a \big]-
{\rm Tr}\big[S_s(x,x)\gamma_\mu \gamma^5 \lambda^a \big] 
{\rm Tr}\big[S_s(0,0)\gamma_\mu \gamma^5 \lambda^a \big]
\bigg).
\label{phi-f1-3}
\end{eqnarray}
Furthermore, in this case, for both terms, it is the chiral-symmetric part of the quark propagator that contributes.  That is
\begin{eqnarray}
    {\rm Tr}\big[S_s(x,x)\gamma_\mu \lambda^a \big]
   & =& \frac{1}{2} {\rm Tr}\big[ \bigg[ S_s(x,x)+ i\gamma^5 S_s(x,x) i \gamma^5 \bigg]\gamma_\mu \lambda^a \big], \\ \nonumber 
    {\rm Tr}\big[S_s(x,x)\gamma_\mu \gamma^5 \lambda^a \big]
   & =& \frac{1}{2} {\rm Tr}\big[ \bigg[ S_s(x,x)+ i\gamma^5 S_s(x,x) i \gamma^5 \bigg]\gamma_\mu \gamma^5 \lambda^a \big].
\end{eqnarray}
Hence, the disconnected contributions in the difference in Eq.~(\ref{phi-f1-3}) are chiral-symmetric, so that even when chiral symmetry is restored, these terms will not vanish.  Although disconnect contributions are expected to be relatively smaller than those from SU(3) breaking effects as discussed in the last case in section 2, they are not non-trivial as discussed in ref.\cite{Lee:2019tvt,Lee:2023ofg}.  Furthermore, in medium, they could be larger.

\section{A case for $K_1$ and $K^*$ in medium}

Therefore, to isolate and observe the effects of chiral symmetry restoration, it is essential to observe both particles that are chiral partners. As mentioned earlier, it is particularly important to study mesons with smaller vacuum widths.
The $\phi,\omega$ and $f_1$'s have small widths and thus are all interesting states to study.  Measuring their medium mass shift will provide important hint on the generation of hadron masses.   However, to study chiral symmetry restoration, it is crucial to also investigate chiral partners.  As shown in Table \ref{t2}, among the chiral parter pairs $(\rho, a_1)$ or $(K^*,K_1)$, only the latter pair has vacuum widths small enough to be realistically measurable. 

\begin{table}[t] 
\centering
\begin{tabular}{ccc|ccc}
\hline
$J^{PC}=1^{--}$ & Mass & Width & $J^{PC}=1^{++}$ & Mass & Width \\
\hline
$\rho$ & 770 & 150 & $a_1$ & 1260 &  250 -600 \\
\hline
$\omega$ & 782 & 8.49 & $f_1$ & 1285 &  24.2 \\
\hline
$\phi$ & 1020 & 4.27 & $f_1$ & 1420 &  54.9 \\
\hline
$K^*(1^-)$ & 892 & 50.3 & $K_1(1^+)$ & 1270 &  90 \\
\hline
$K^*(1^-)$ & 1410 & 236 & $K_1(1^+)$ & 1400 &  174 \\
\hline
\end{tabular}
\caption{Physical parameters of the vector and axial vector mesons. Units are in MeV$/c^2$.}\label{t2}
\end{table}

\begin{table}[b] 
\caption{Dominant hadronic decay channels of $K^*$ and $K_1$ mesons.\label{t1}}
\centering
\begin{tabular}{cc|cc}
\hline
$1^{-}$ & Decay Mode &   $1^{+}$ & Decay mode  \\
\hline
$K^*(892)$ & $K \pi$ (100\%) & $K_1(1270)$ & $K \rho$ (42\%) \\
 & &    &  $K^* \pi$ (16\%) \\
\hline
\end{tabular}
\end{table}

The decay channel that is dominant for the $K^*$ and $K_1$ mesons are given in Table \ref{t1}.
There are different charge states for the $K^*$ and $K_1$. 
The chiral partners are between the same charge states.  If the baryon density of the medium is not zero, the different charged states will respond differently irrespective of chiral symmetry restoration.  In such an environement, it is important to compare $K^*$ and $K_1$ with the same flavor states.  For states with an s-quark, the decay modes are given as follows. For the $K_1$,
\begin{align}
K_1^- \to \begin{cases}
\rho^0 K^-  \cr
\rho^- \bar{K}^0  \cr  
\pi^0 K^{*-}  \cr
\pi^- \bar{K}^{* 0}  
\end{cases}
, ~~~~
\bar{K}_1^0 \to \begin{cases} 
\rho^+ K^-  \cr
\rho^0 \bar{K}^0 \cr   
\pi^+ K^{*-}  \cr
\pi^0 \bar{K}^{* 0}  \end{cases}. 
\label{k1-decay}
\end{align}
It should be noted that the $\rho$ and $K^*$ decay into $\pi \pi$ and $K\pi$, respectively. Therefore all the decays can be seen in the $\pi\pi K$ final states. The $K_1$ are indeed seen in  pp event at low multiplicity and are being studied at higher multiplicities at LHC\cite{Sanghoon24} .

For the corresponding $K^{*-}$, it is seen through the following decays.
\begin{align}
K^{*-} \to \begin{cases} 
\pi^0 K^{-}  \cr
\pi^- \bar{K}^{ 0} \end{cases}  
, ~~~~
\bar{K}^{*0} \to \begin{cases} 
\pi^+ K^{-}  \cr
\pi^0 \bar{K}^{ 0} \end{cases} .
\label{kstar-decay}
\end{align}

\subsection{$K_1$ and $K^*$ in nuclear matter}

The production of both of these particle on a nuclear target can be achieved using the Kaon beam at the JPARC facility. For example, if a $K^-$ beam is used, the produced these particles and their final states are given in Eq.~(\ref{k1-decay}) and Eq.~(\ref{kstar-decay}) for $K_1$ and $K^*$, respectively.

It is important to measure both particles with small velocity because the spin-1 particles will respond differently whether the spin is alinged parralell or transverse to its motion with respect to the medium at rest.  This is so because the mass of the transverse or longitudinal modes will shift in the opposite direction\cite{Lee:1997zta,Kim:2019ybi}.  
This effect is dominated by kinematical effect and is not related to chiral symmetry restoration in the medium\cite{Lee:2023ofg}. 
One can experimentally separate the transverse and longitudinal modes through the angular dependence of the two-body decay of these particles\cite{Park:2022ayr,Park:2024vga}. Then one can experimentally identify the momentum independent mass shifts. 
Similar production of both the $K^*$ and $K_1$ can be achieved by a pion beam at GSI\cite{Paryev:2020ivs}. 

\subsection{$K_1$ and $K^*$ from heavy ion collisions}

The degeneracy between  $K^*$ and $K_1$ mass when chiral symmetry is restored can also be probed in a relativistic heavy ion collision\cite{Sung:2021myr,Sung:2023oks}.  
This is because the initial temperature in an ultra-relativistic heavy-ion collision is expected to be above the transition point to the quark-gluon plasma phase, where chiral symmetry is expected to be restored. As the system cools, hadrons will form at around 156 MeV, as estimated by the statistical hadronization model (SHM) through the observed  production of hadron abundances\cite{Stachel:2013zma}. At this temperature, however, the chiral order parameter is still quite small \cite{Ding:2013lfa}, and since the mass difference depends on the chiral order parameter, the masses of chiral partners will be similar.
Since $K*$ and $K_1$  have the same strangeness, their production ratios will be similar at the hadronization point. 

One must still consider the possible changes in particle numbers as the system passes through the hadronic phase \cite{Cho:2015qca}. This is known as the hadronic rescattering effect, which typically depends on the vacuum width of the hadron in question. In a central collision, which has a longer hadronic lifetime, the initial 
$K^1/K^*$ production ratio—close to 1 at the hadronization point—will not be preserved due to hadronic rescattering, which significantly reduces the ratio as the hadrons acquire their vacuum masses at the system evolves in the hadronic phase.
If the vacuum masses are used in the SHM, the expected particle ratio between $K_1$ and $K^*$ will be very small due to the much larger mass of $K_1$ in the vacuum. However, in peripheral collisions, where the collision region remains dominated by initial temperatures above the transition point, the hadronic lifetime will be shorter, and the anomalously larger initial particle ratio will be visible. 
Therefore measuring the production of both the $K_1$ and $K^*$ from heavy ion collision in both central and peripheral collisions, and comparing the observed production ratios to those obtained in the SHM with vacuum masses, will provide a clear signature of chiral symmetry restoration in heavy ion collisions\cite{Sung:2021myr,Sung:2023oks}.

\section{Summary and Conclusions}\label{sec:sumandconc}

In this work, we have tried to clarify the concept of chiral partners. For a vector meson, the isospin-zero and hypercharge-zero state in the flavor octet mixes with the flavor singlet state. Since the flavor singlet vector meson does not have a chiral partner, the mixed $\omega$ and $\phi$  mesons will not have chiral partners. This means that even when chiral symmetry is restored, these mesons will not become degenerate with their corresponding parity partners.  
On the other hand, the $K_1$ and $K^*$ 
mesons are chiral partners, and both have widths smaller than 100 MeV. Therefore, we emphasize that studying these mesons in environments where chiral symmetry is restored is particularly important for understanding the effect of chiral symmetry restoration on chiral partners and their masses.

It is still interesting and important to measure the $\omega$, $\phi$ and $f_1$ mesons in medium as well as they have small width so that one can reconstruct their in medium masses. If such measurement can be made, it will provide important clue to understand the generation of hadron masses in general.

\section*{Acknowledgments}
This work was supported by the Korea National Research Foundation under grant No. 
2023R1A2C300302312 and No. 2023K2A9A1A0609492411.

\end{document}